\documentclass{epl}

\title{Spontaneous magnetization and Hall effect in superconductors
with broken time-reversal symmetry \\}
\author{Baruch Horovitz and Anatoly Golub}
\institute{Department of Physics, Ben-Gurion University of the Negev,
Beer-Sheva 84105, Israel}

\begin{document}

\maketitle

\begin{abstract}
Broken time reversal symmetry (BTRS) in d wave superconductors is
studied and is shown to yield current carrying surface states.
 The corresponding spontaneous
magnetization $\Phi$ is temperature independent near the critical
temperature $T_c$ for weak BTRS, in accord with recent data.  
For strong BTRS and thin films we expect a temperature dependent 
$\Phi$ with a paramagnetic anomaly near $T_C$.
The Hall conductance is found to vanish at zero wavevector $q$ and
finite frequency $\omega$, however at finite $q, \omega$ it has an
unusual structure.
\end{abstract}

Recent data on the high $T_c$ superconductor $YBa_2Cu_3O_x$ (YBCO)
has supported the presence of broken time reversal symmetry (BTRS)
\cite{covington,deutscher,polturak}.  A sensitive probe of BTRS are 
Andreev surface states. For a d wave with time reversal symmetry bound 
states at zero energy are expected for a surface parallel to the 
nodes (a (110) surface in YBCO). When BTRS is present, by either a 
complex order parameter or by an external magnetic field, the bound 
states shift to a finite energy. Indeed tunneling data usually shows 
a zero bias peak which splits in an applied field; the splitting is 
nonlinear in the magnetic field, indicating a proximity to a BTRS state 
\cite{deutscher,alff}. In fact, in some samples tunnelling data
shows a splitting even without an external field
\cite{covington,deutscher}, consistent with BTRS; the splitting
increases with increasing overdoping \cite{deutscher,sharoni},

Further support for a spontaneous BTRS state are spontaneous magnetization
data as observed in YBCO
\cite{polturak}, setting in abruptly at $T_c$ and being almost
temperature ($T$) independent below $T_c$.  The phenomenon has been
attributed to either a $d_{x^2-y^2}+id_{xy}$ state or to formation
of $\pi$ junctions.  No microscopic reason was given, however, for
the spontaneous magnetization being independent of both $T$ and of
 film thickness \cite{polturak}.

It has been shown theoretically that BTRS can occur locally in a
$d_{x^2-y^2}$ superconductor near certain surfaces
\cite{sigrist1,matsumoto,rainer} leading to surface currents.  The
onset of such BTRS is expected to be below $T_c$ and therefore
does not correspond to the spontaneous magnetization data
\cite{polturak}.  We note that in response to an external 
magnetic field the surface states are paramagnetic and compete with Meissner 
currents. This effect has been proposed to account for a minimum in 
the magnetic penetration length \cite{walter}. In fact, it was 
proposed that this paramagnetic effect leads to a spontaneous currents 
and BTRS in a pure $d_{x^2-y^2}$ state \cite{higashitani,barash}. 
The onset of this BTRS is much below $T_c$ \cite{barash} and 
therefore does not correspond to the data \cite{polturak}.

Of further theoretical interest is the relation
of the BTRS state to quantum Hall systems with a variety of Hall
effects \cite{goryo,horovitz,read,goryo2}.  In particular a finite charge
hall conductance has been suggested \cite{goryo}, though this has
been questioned \cite{read}.

In the present work we assume that BTRS is a bulk property, i.e.
both components of an order parameter $d_{x^2-y^2}+id_{xy}$ set in
at $T_c$; this is a plausible scenario which can possibly account 
for the data. In previous works \cite{sigrist1,matsumoto,rainer}
surface states appear due to a surface induced $id_{xy}$
component.  We find, however, the less anticipated result that the
{\em bulk} state $d_{x^2-y^2}+id_{xy}$ leads to surface states
with finite surface current densities.  The latter situation was found in
the bulk p wave state \cite{sigrist2} and for the total current was inferred from 
topological considerations \cite{volovik}.  We then evaluate the
spontaneous magnetization and show that it is dominated by $(100)$
surfaces; for thin films it increases with the ratio
 $\Delta '/\Delta$ ($\Delta$ and $\Delta '$ are the amplitudes of
$d_{x^2-y^2}$ and $d_{xy}$, respectively) while for thick films it has
a maximum at $\lambda/\xi ' \approx 1$ where $\xi$, $\xi '$ are
the coherence lengths associated with $\Delta$, $\Delta'$, 
respectively ($\xi'=$Fermi velocity$/\Delta'$, similarly for $\xi$)
and $\lambda$ is the magnetic penetration length. The maximum for 
YBCO is at $\Delta '/\Delta \approx \xi/\lambda\approx 0.01$. 
Throughout we assume an exterme type II superconductor, i.e. $\xi \ll 
\lambda$, while $\xi '/\lambda$ is arbitrary. We show that for weak
BTRS, $\lambda/\xi'<1$, the spontaneous magnetization is $T$ and
thickness independent, while for strong BTRS thickness and $T$
dependence may occur, as well as a transition to a paramagnetic state 
close to $T_c$. For the sample of Ref
\cite{polturak} we estimate $\Delta '/\Delta \approx
10^{-4}$, i.e. weak BTRS. We also derive the effective action and
identify the Hall coefficient which has an unusual wavevector and frequency
dependence.

First we demonstrate the existence of surface states.  Consider a
$d_{x^2-y^2}+id_{xy}$ state where the order parameter is
\begin{equation}
 \Delta ({\hat p}_x,{\hat p}_y)=\Delta '{\hat p}_x{\hat p}_y/k_F^2
 +i\Delta({\hat p}_x^2-{\hat p}_y^2)/k_F^2
 \end{equation}
where ${\hat {\bf p}}=-i\hbar {\mbox{\boldmath $\nabla$}}$ is the
momentum operator and $k_F$ is the Fermi momentum.  We consider a
vacuum-superconductor boundary at x=0, and assume for now that
$\Delta, \Delta '$ are constants at $x>0$ and vanish at $x<0$.  For
$\Delta \gg \Delta '$ this corresponds to a $(100)$ surface; to
 describe a $(110)$ surface $\Delta$ and $\Delta '$ need to be
 interchanged.  The
electron-hole wavefunctions $u(x,k_y)\exp (ik_yy)$, $v(x,k_y)\exp
(ik_yy)$ with mass $m$ satisfy the Bogoliubov de-Gennes equations
\begin{eqnarray}
(2m)^{-1}(-k_F^2-d^2/dx^2)u(x,k_y) +\Delta ({\hat p}_x,k_y)v(x,k_y)
&=&
\epsilon u(x,k_y) \nonumber\\
\Delta ^* ({\hat
p}_x,k_y)u(x,k_y)+(2m)^{-1}(k_F^2+d^2/dx^2)v(x,k_y)&=&
\epsilon v(x,k_y) \label{BdG}
 \end{eqnarray}
The decaying eigenfunctions have the form $\sim \exp(\pm
ikx-x|k_y|/\xi 'k_F)$ with $k=|k|$ and $\xi '=k_F/m\Delta '$.
 Specular reflection at the surface requires a
 superposition of $\pm k$ states which vanish at the surface.  This
 yields the eigenvalue equation
\begin{equation}
\frac{i\epsilon + \sqrt{|\Delta(+ k,k_{y})|^{2}-\epsilon^{2}}}
{-i\epsilon + \sqrt{|\Delta(- k,k_{y})|^{2}-\epsilon^{2}}}=
-\frac{\Delta(+ k,k_y)}{\Delta(- k,k_y)} \label{eigen}
\end{equation}
Its solutions are $\epsilon =-{\mbox sign}
     (k_y)\Delta(k^2-k_y^2)/k_F^2$.  In terms of the incidence angle
     $\zeta$, $k_y=k_F\sin \zeta$, $k=k_F \cos \zeta$,
     the allowed positive eigenvalues
      are $\epsilon =\Delta \cos (2\zeta)$ for $-\pi /4<
     \zeta <0$ and $\epsilon =-\Delta \cos (2\zeta)$ for
$\pi/4<\zeta <\pi/2$ (see inset of Fig.  1).  We note that self
consistency would imply that $\Delta '=0$ at $x=0$ \cite{rainer};
the eigenfunctions would then be $\sim \exp[-\int_0^x\Delta
'(x')dx'|\sin \zeta|/v_F]$, resulting in a qualitatively similar
dependence on $\xi '$.  The dominant order parameter $\Delta$ is
finite at the $(100)$ boundary \cite{rainer}, hence we expect our
results to be quantitatively correct.

In presence of a vector potential $A_y(x)$ the spectrum 
$\epsilon$ is Doppler shifted by $-(e/mc)k_yA_y$.
For the expectation value of the current density in the y
direction and the charge density we obtain

\begin{eqnarray}
\frac{4\pi}{c}j_{edge}(x)&=& \frac{4\phi _0}{\pi \xi '\lambda
_0^2}\int_0^{\pi /2}d\zeta \cos \zeta \sin ^2\zeta e^{-2x\sin
\zeta /\xi '} \tanh(\frac{\Delta \cos 2\zeta+
(e/c)v_F\sin\zeta A_y(x)}{2T})\nonumber\\
n_{edge}(x)&=& \frac{ek_F}{\pi d \xi '}\int_0^{\pi /2}d\zeta \cos
\zeta \sin \zeta e^{-2x\sin \zeta /\xi '} \label{edge}
\end{eqnarray}
where $\lambda_0=(mc^2d/2k_F^2e^2)^{1/2}=\lambda(T=0)$ and $d$ is
the spacing between layers, converting the current per layer of
the states in Eq.  (\ref {BdG}) to a current density.  Note that for
either $\Delta =0$ or $\Delta ' =0$ all angles $\zeta$ are allowed in the
solution of Eq.  (\ref{eigen}) and then the current density vanishes.  This
demonstrates then that BTRS leads to current carrying surface
states.  We note also that the integrated current 
$\int_0^{\infty}j_{edge}(x)dx$
vanishes, unlike the p wave case \cite{sigrist2}.

\begin{figure}
\onefigure[scale=0.7]{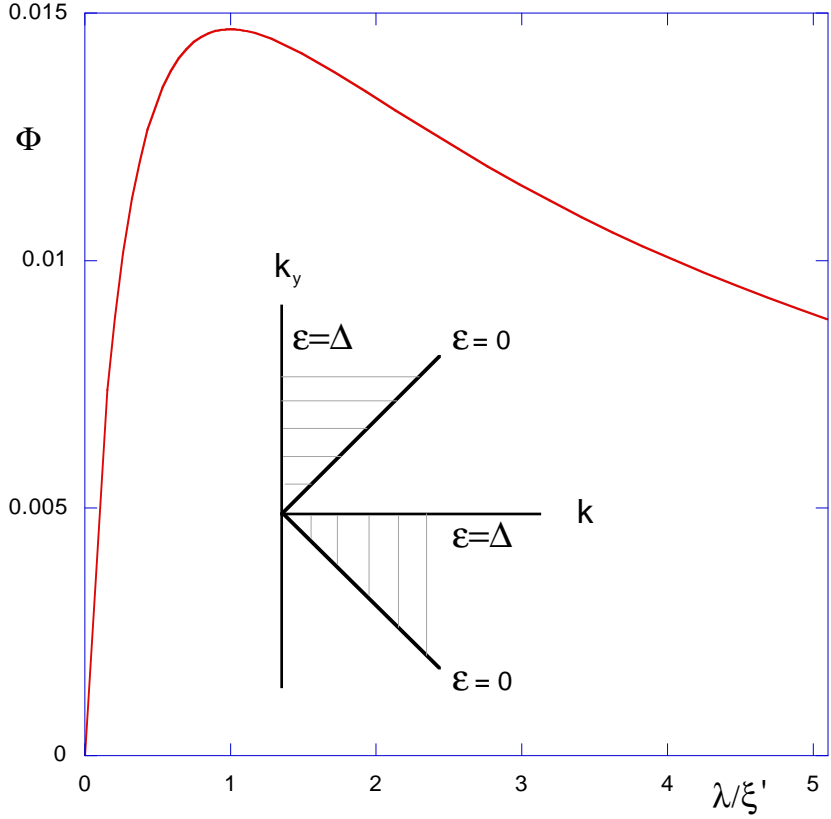}
\caption{Spontaneous flux for a $(100)$ boundary in thick films
 ($\xi'<{\bar d}$)
 in units of $2\phi _0L_y\lambda \Delta/\pi \lambda
_0^2T_c$.  The hatched areas in the inset show the allowed directions
 ($k,k_y$) of bound
states.  Their spectra spans the range $\epsilon =0$ (diagonal lines)
up to $\epsilon = \Delta$ as shown.}
\end{figure}

The response to $j_{edge}(x)$ includes London terms as well as a
BTRS induced Chern-Simon term.  As shown below, the Chern-Simon
term is weaker at $T\rightarrow T_c$ so that London's equation
with $j_{edge}(x)$ as a source term is valid,
\begin{equation}
    -\nabla ^2 A_y(x)=[-(1/\lambda^2)A_y(x)
    +(4\pi /c)j_{edge}(x)]\theta (x) \label{London}
    \end{equation}
    where $\theta (x)$ is a step function.
We consider first solutions where the Doppler shift $\sim A_y$ on the 
right hand side of Eq~(\ref{edge}) is neglected. 
    This is solved by the Greens' function $G(x,x')=-(\lambda
    /2)[\exp (-|x-x'|/\lambda)+\exp (-|x+x'|/\lambda)]$ resulting at
    $T\rightarrow T_c$ in the solution
\begin{equation}
A_y(0)=(2\phi _0\lambda \Delta/\pi \lambda _0^2T_c)\int
_0^{\pi/2}d\zeta \cos \zeta \sin ^2\zeta \cos 2\zeta (2\sin \zeta
+\xi '/\lambda)^{-1}
\end{equation}
The total spontaneous flux is $\Phi =A_y(0)L_y$ where $L_y$ is
the length of the boundary.  The ratio ${\tilde \Phi}=-\Phi /
(2\phi _0L_y\lambda \Delta/\pi \lambda _0^2T_c)$ is shown in Fig.
1.  It varies between $\xi'/12\lambda$ at $\lambda \gg \xi '$ to
$\lambda /15\xi '$ at $\lambda \ll \xi '$ with a maximum of
$0.014$ at $\lambda \approx \xi '$.  For a $(110)$ surface ${\tilde
\Phi}=\xi\Delta '/12\lambda \Delta$, much smaller than for a
$(100)$ surface.  We note that $\Phi$ is $T$ independent up to
$T_c$ since the product $\lambda \Delta$ is finite at
$T\rightarrow 0$, consistent with the spontaneous magnetization
data \cite{polturak}; more details on the data follow in the
discussion.

The above refers to thick films with thickness ${\bar d}>\xi'$.
For thin films the right hand side of Eq.  (\ref{London}) is
multiplied by ${\bar d}\delta (z)$.  For $A_y(x)\equiv A_y(x,
z=0)$ this yields
\begin{equation}
A_y(x)-A_y(0)=\int _0^{\infty}dx'\ln |(x-x')/x'|[-1/\lambda^2)A_y(x')
    +(4\pi /c)j_{edge}(x')]{\bar d}
    \end{equation}
which implies a slow (non-exponential) decay of $A_y(x)$.  To avoid
divergence of $dA_y(x)/dx|_{x=0}$ the relation $A_y(0)=\lambda ^2(4\pi
/c)j_{edge}(0)$ must hold.  This, interestingly, yields for $\Phi$
the previous result of the $\lambda \ll \xi '$ case, i.e. ${\tilde
\Phi}=\lambda /15\xi '$; in Fig.  1 this is the tangent line to the
thick film curve at the origin.
 Hence we can define two regimes: Weak BTRS
with $\lambda/\xi'<1$ where the spontaneous flux is $T$ and ${\bar d}$
 independent, and strong BTRS with $\lambda/\xi'>1$ where film
thickness matters.  In the latter case $T$ dependence is induced as
$\xi'<{\bar d}$ changes to the thin film case $\xi'>{\bar d}$ as
$T\rightarrow T_c$.

We now reconsider the effect of the Doppler shift. Near $T_c$ the 
effect is linear in $A_y$ and can be incorporated as an effective 
$\lambda_{eff}$ replacing $\lambda$ in the results above, where
\begin{equation}
\frac{1}{\lambda _{eff}^2}\approx\frac{1}{\lambda^2}-\frac{\xi_0}
{\lambda_0^2\max (\xi',\lambda)} \,\, .
\end{equation}
Hence at a temperature $T_c'$ $\lambda_{eff}$ changes sign, where 
$(T_c-T_c')/T_c\approx[\xi/\max (\xi',\lambda)]^2$, i.e. less then $10^{-4}$ 
for YBCO. The effect of the Doppler shift was previously considered 
only at low temperatures where on the (110) surface paramagnetism and spontaneous 
surface currents appear even when $\Delta'=0$ at $\sim 
(\xi_0/\lambda_0)T_c\ll T_c$ 
\cite{higashitani,barash}. For thin films $\max (\xi',\lambda)$ is 
replaced by $\xi'$ and $(T_c-T_c')/T_c\approx(\xi/\xi')^2$ also for strong 
BTRS where $T_c-T_c'$ is enhanced. Hence a transition from 
paramagnetic ($T>T_c'$) to diamagnetic response ($T<T_c'$) can be observed at 
$T_c'$ and the magnetization ${\tilde \Phi}=(\lambda /15\xi ')
(\lambda_{eff}/\lambda)^2$ is peaked and changes sign at $T_c'$. The 
effect is more pronounced for the (110) surface, however, the 
condition for a thin film ${\bar d}<\xi$ is more difficult to achieve.

Next we consider the effective action of a bulk
$d_{x^2-y^2}+id_{xy}$ superconductor.  In terms of the Nambu
spinors $\psi^{\dagger}({\bf r})=[u^*({\bf r}),u^*({\bf r})]$, the 
superconducting phase $\theta ({\bf r})$ and
Pauli matrices $\tau_i$, the transformation $\psi({\bf
r})\rightarrow exp[i\tau_{3} \theta ({\bf r})/2] \psi({\bf r})$
yields the off-diagonal Hamiltonian $\int d^2r \psi^{\dagger}({\bf
r})h_{\Delta}\psi ({\bf r})$ where
\begin{equation}
 h_{\Delta }=-[\Delta
(-\partial_{x}^{2}+\partial_{y}^{2})\tau_{1}+\Delta'
\partial_{x}\partial_{y}\tau_{2}]/k_{F}^{2} \,
\end{equation}
and we neglect terms with $\nabla\theta<< k_{F}$.  The action in
presence of
electromagnetic potentials ${\bf A}, \varphi $ is then
\begin{eqnarray}
S&=&\int d^2rdt \psi^{\dagger}(i\partial_{t}-\tau_{3}
\epsilon(\hat{p})-h_{\Delta }
-\Sigma)\psi \nonumber\\
\Sigma &=&\tau_{3}(a_0+{\bf a}^2/2m) + {\bf a}\cdot {\bf p}/m- i
{\mbox{\boldmath $\nabla$}}\cdot{\bf a}/2m
\end{eqnarray}
where $\epsilon(\hat{p})=(\hat{p}^2-k_F^2)/2m$ and we introduce
the gauge invariant potentials ${\bf
a}=\frac{1}{2}{\mbox{\boldmath $\nabla$}}\theta-e{\bf A}$ and
$a_{0}= \frac{1}{2}\frac{\partial}{\partial t}\theta-e\varphi$.
Expansion to 2nd order in ${\bf a},a_0$ leads to the effective
action
\begin{equation}
S_{eff}=\int \frac{d^{2}q d\omega} {(2\pi)^3}P_{\mu\nu}({\bf
q},\omega) a_{\mu}({\bf q},\omega ))a_{\nu}(-{\bf q},\omega))
\label{Seff}
\end{equation}
At $T=0$ and ${\bf q},\omega \rightarrow 0$ we obtain $P_{00}=N_0$
(density of states which is $N_0=m/2\pi$ in two dimensions),
$P_{ij}=-N_0c_s^2$ where $c_s^2=v_F/\sqrt{2}$, while $P_{0j}(q)=
i{\mbox sign}(\Delta\Delta')\epsilon_{0ij}q_{i}/(4\pi)$ and
$\epsilon_{0ij}$ is the antisymmetric unit tensor.  The latter term
reflects BTRS and is derived for $\Delta '\ll \Delta$.

Integrating out the phase $\theta$ we obtain the effective action in
terms of the electromagnetic potentials ${\bf A},\varphi$
\begin{eqnarray}
S_{eff}\{{\bf A},\varphi\}=e^2 \int \frac{d^{2}q
d\omega}{(2\pi)^3}\{&& \frac{c_{s}^2 {\bf q}^2}{c_{s}^2 {\bf
q}^2-\omega^2} [P_{00}|\varphi({\bf q},\omega)|^2- \frac{i}{4\pi}
\epsilon_{0ij}q_{i}\varphi({\bf
q},\omega)A_{j}(-{\bf q},-\omega) \nonumber\\
&&+O(\omega ^2|{\bf A}|^2)]-P_{00}(\frac{c_{s}}{c})^{2}|{\bf
A}({\bf q},\omega)|^2\}
\end {eqnarray}

The total electromagnetic action includes also the Maxwell field
part $S_{M}=\int d^2rdt(\vec{E}^2-\vec{H}^2)/8\pi$.  E.g., for
$\omega\neq 0$, $q\rightarrow 0$ the propagator for $\varphi$
yields the plasmon mode at $\omega_p=c/\lambda_0$.  The Hall
current $J_y$ is identified by a functional derivative with
respect to $A_x$ leading to the Hall coefficient
\begin{equation}
\sigma_{xy}({\bf q},\omega)={\mbox sign}(\Delta\Delta')\frac{e^2}{4\pi
\hbar}
\frac{c_{s}^{2}q^2}{c_{s}^{2}q^{2}-\omega^{2}} \label{xy}
\end{equation}
Transport is defined by taking the $q\rightarrow 0$ limit first,
i.e. $\sigma_{xy}=0$.  Hence the conventional Hall coefficient
vanishes, as expected from Galilean invariance \cite{read}.  A
limit in which $\omega \rightarrow 0$ is taken first yields a
quantized "static" conductance $e^2/2h$ which was argued to
correspond to $\sigma_{xy}\neq 0$ in presence of a boundary
\cite{goryo}.  In absence of an external magnetic field, and given
a spontaneous magnetization decaying in the bulk (as confirmed below),
Amp{\'e}re's law
yields zero total current, hence $\sigma_{xy}=0$; this is valid
also with a boundary and external electric field.
It is intriguing, however, that $\sigma_{xy}({\bf q},\omega)$ has
a nontrivial structure and space resolved measurement of a Hall
current could then probe the full Eq.  (\ref{xy}). We note that a 
result similar to Eq. (\ref{xy}) was obtained for superfluid $^3$He 
\cite{goryo2}.

We proceed to derive the effective action in presence of a
boundary and at finite $T$.  Special care is needed for the
Chern-Simon coefficient $P_{0j}$ which now nonlocal due to
specular reflection at the boundary.  After integrating out
$\theta$ we obtain the form (considering only the $\omega=0$ term)
\begin{eqnarray}
 S_b\{{\bf A},\varphi\} & = & e^2 \int dr[P_{00}(\varphi^2({\bf r})-
( \frac{c_{s}}{c})^{2}{\bf A}^2({\bf r}))
+b_1({\bf r}) \varphi({\bf r}) A_{y}({\bf r})
+b_2({\bf r}) \varphi({\bf r}) \frac{ \partial A_{y}({\bf
r})}{\partial x}]
\end{eqnarray}
Variation of $S+S_M$ leads to a generalization of Eq.  (\ref
{London}) in which the equations for ${\bf A}$ and $\varphi$ are
coupled,
\begin{eqnarray}
(\frac{1}{\lambda_{d}^{2}}-\frac{\partial^{2}}{\partial
x^{2}})\varphi-\frac{4\pi e^2}{c\hbar}b_{2}
\frac{\partial A_{y}}{\partial x}-\frac{4\pi e^2}{c\hbar}b_{1}
A_{y}&=& 4\pi e n_{edge}(x) \label{nedge}\\
(\frac{1}{\lambda^{2}}-\frac{\partial^{2}}{\partial
x^{2}})A_{y}-\frac{4\pi e^2}{c\hbar}b_{2}\frac{\partial
\varphi}{\partial x}+\frac{4\pi e^2}{c\hbar}(b_{1}-\partial_{x}
b_{2}) \varphi &=&-\frac{4\pi }{c}j_{edge}(x)\label{jedge}
\end{eqnarray}
where $\lambda_d=(8\pi e^2N_0)^{-1/2}$ is the Debye screening
length.  For $T\rightarrow T_{c}$ we obtain $\lambda\approx
\lambda_0(1-T/T_c)^{-1/2}$, $b_1({\bf r})=
0.11(\Delta \Delta'/T_{c}^{2})\frac{d}{dx}ln[\Delta
\Delta']/2hcd$ and $b_2({\bf r})= 0.21(\Delta
\Delta'/T_{c}^{2})/2hcd$.

An external electric field leads to $\varphi \sim \exp
(-x/\lambda_d)$, hence the magnetization remains localized even in
presence of such a field and
space integration of Eq.  (\ref {jedge}) (i.e. Amp{\'e}re's law)
yields a zero total current, i.e. $\sigma_{xy}=0$.  The Chern-Simon
term, however, affects the spontaneous magnetization, leading to
an additional flux $\sim (\Delta '/\Delta) (1-T/T_c)$ which
vanishes at $T\rightarrow T_c$.

We consider now in more detail the experimental data on the
spontaneous magnetization \cite{polturak}.  The data shows that for
a YBCO disc with a perimeter of $L_y\approx 2$cm the spontaneous
magnetization is temperature independent in the range 80-89K and is 
also thickness independent in the range 30-300nm with a value of  
$\approx$37$\phi_0$.  Taking $\lambda \Delta \approx
\lambda _0\Delta _0$, their $T=0$ value, and typical YBCO
parameters we find ${\tilde \Phi }\approx 10^{-3}$.  Fig.  1 implies
that the limit $\xi '>\lambda$ applies i.e. weak BTRS; hence for
either thick or thin films we estimate
 $\lambda/\xi '\approx 10^{-2}$ or
$\Delta'/\Delta\approx 10^{-4}$.  We propose therefore that
increasing the ratio $\Delta'/\Delta$, e.g. by using overdoped
YBCO \cite{deutscher}, one can enhance the spontaneous
magnetization up to a maximum of $\approx 10^3\phi_0$ when
$\Delta'/\Delta \approx 0.01$ within the thick
film regime.

For strong BTRS, $\lambda/\xi'>1$, the film thickness
matters, i.e. we expect a temperature
dependence due to the crossover from thick to thin film regimes at 
${\bar d}\approx\xi'$ as
$T\rightarrow T_c$.  For thin films (${\bar d}<\xi'<\lambda)$ we
obtain ${\tilde \Phi}=\lambda /12\xi'$, i.e. for
YBCO the total flux can reach $10^5\Delta'/\Delta\phi_0$ per cm of
boundary, much higher than thick film values. The situation of a 
strong BTRS with thin films is interesting also as being the most 
likely one to show the 
paramagnetic anomaly at $T_c'\approx T_c[1-(\xi/\xi')^2]$.

In conclusion, we have shown that surface states of a
$d_{x^2-y^2}+id_{xy}$ superconductor lead to spontaneous
magnetization which is $T$ independent and thickness independent
for weak BTRS, $\lambda/\xi'<1$, in accord with the
data \cite{polturak}.  For strong BTRS, $\lambda/\xi'>1$,
 as expected in overdoped YBCO \cite{deutscher} , a
crossover from thick to thin film behavior can lead to $T$ and
thickness dependence, as well as to an observable paramagnetic anomaly 
near $T_c$.  We also find that the Hall conductance has
an unusual $\sigma_{xy}({\bf q},\omega)$ dependence, though its
conventional transport value vanishes.

We thank E. Polturak and A. Stern for valuable discussions.  This
research was supported by THE ISRAEL SCIENCE FOUNDATION founded by
the Israel Academy of Sciences and Humanities.



\end{document}